\documentclass[english]{article}
\usepackage[T1]{fontenc}
\usepackage[latin9]{inputenc}
\usepackage{geometry}
\usepackage{verbatim}
\usepackage{amsmath}
\usepackage{amsthm}
\usepackage{amssymb}
\usepackage{graphicx}
\usepackage{setspace}
\usepackage{hyperref}
\makeatletter

\newcommand{\lyxdot}{.}

\numberwithin{equation}{section}
\numberwithin{figure}{section}
\theoremstyle{plain}
\newtheorem{thm}{\protect\theoremname}
\theoremstyle{plain}
\newtheorem{prop}[thm]{\protect\propositionname}

\usepackage{dsfont}

\@ifundefined{showcaptionsetup}{}{%
 \PassOptionsToPackage{caption=false}{subfig}}
\usepackage{subfig}
\makeatother

\usepackage{babel}
\providecommand{\propositionname}{Proposition}
\providecommand{\theoremname}{Theorem}

\begin{document}

\title{Quantitative earnings enhancement from share buybacks\thanks{Research sponsored by The Witness Corporation (www.thewitnesscorporation.com)}}

\author{Lawrence Middleton\thanks{Department of Statistics, University of Oxford},
James Dodd\thanks{St. Cross College, University of Oxford}, Graham
Baird\thanks{Department of Mathematics, University of Oxford}}
\maketitle
\begin{abstract}
This paper aims to explore the mechanical effect of a company's share
repurchase on earnings per share (EPS). In particular, while a share
repurchase scheme will reduce the overall number of shares, suggesting
that the EPS may increase, clearly the expenditure will reduce the
net earnings of a company, introducing a trade-off between these competing
effects. We first of all review accretive share repurchases, then
characterise the increase in EPS as a function of price paid by the
company. Subsequently, we analyse and quantify the estimated difference
in earnings growth between a company's natural growth in the absence
of buyback scheme to that with its earnings altered as a result of
the buybacks. We conclude with an examination of the effect of share
repurchases in two cases studies in the US stock-market.
Accompanying code can be found at \url{https://github.com/particlemontecarlo/quantifying_eps_buybacks}.
\end{abstract}

\section{Introduction}

\subsection{Share repurchases background}

Share repurchases provide a popular means for companies to return
cash to their shareholders as an alternative to stock dividends \cite{jagannathan2000financial,pettit2001share}.
Popularised in the 80s, following a change in regulations governing
open market share repurchase schemes, they became a means for managers
to alter earnings per share via financial engineering \cite{grullon2000we,fox2018thebig}
thereby having a knock-on effect on the share price. Since, repurchase
programs have generally increased in US stocks \cite{gruber2018corporate},
with the value of buybacks of S\&P 500 companies representing a significant
portion of US gross domestic private investment. In general, the precise
reasons for a company to buybacks its shares may vary \cite{dobbs2005thevalue},
however efforts have shown that companies may use them as an earnings
management device, with evidence to suggest the intention of aiming
to inflate the earnings per share  \cite{hribar2006stock}. Indeed,
while share repurchases inevitably reduce the number of shares outstanding
(thereby reducing the divisor of the earnings per share), such an
action will inevitably shrink a company's asset base thereby potentially
having a negative effect on the earnings per share \cite{bens2003employee}. 

From an investment perspective, on the one hand it may provide a positive
sign that a company's board believes in the future growth in the company
- enough to treat the repurchase as an investment decision \textendash{}
on the other hand such a decision may also reflect that the company
has limited investment opportunity (as well as the aforementioned
motivation to inflate earnings) \cite{grullon2004information}. 

\subsection{Share repurchases in the media}

Recently, buybacks have made headlines due to a `record year' in buybacks
\cite{henderson2019share} and the recent activity following the very
high level of cash reserves held by a number of companies, particularly
in the technology sector, \cite{waters2019share,waters2018five} (curtailing
only recently \cite{henderson2019fall}), with similar activity observed
in the aerospace industry \cite{ford2019boeing}. A recent surge in
US repurchases has been observed in part due to Trump's tax reform
thereby freeing up capital \cite{rocco2019us}. The FT article `Negative
interest rates fuel record Japan share buybacks' (May 24, 2016) remarks
the increase buybacks as a result of negative interest rates in Japan.
However, the surge in buyback activity has seen criticism in \cite{grant2019big}
as newly available cash reserves are spent on repurchases benefitting
investors rather than boosting employment or R\&D. Additionally, a
correlation in \cite{rocco2019stock} has been observed between the
increase in sell-offs of single stocks and an increase in purchasing
ETFs related to buybacks.

\begin{figure}
\begin{centering}
\subfloat[Share buybacks on S\&P 500]{\includegraphics[scale=0.4]{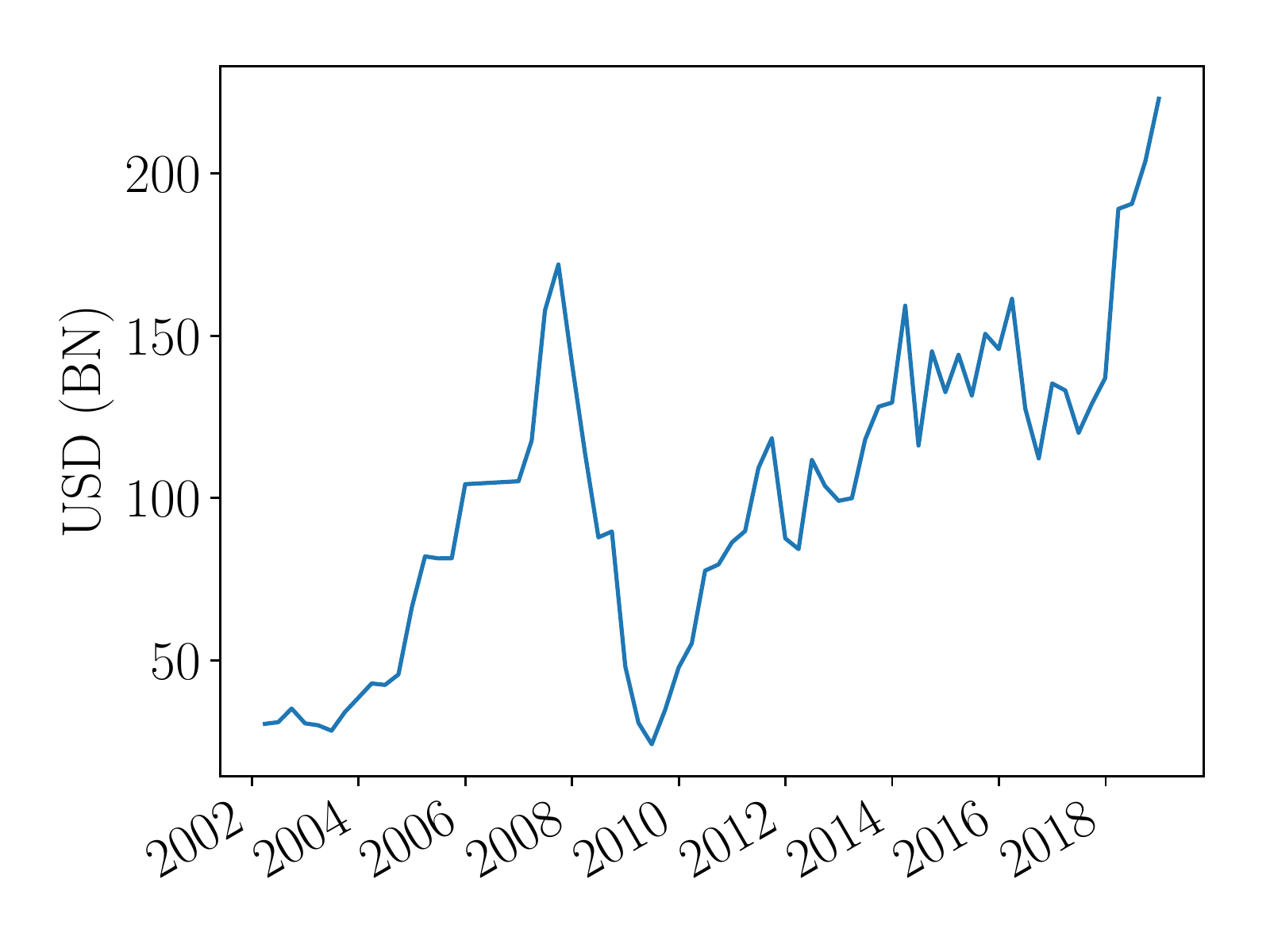}}\subfloat[Google interest index]{\includegraphics[scale=0.4]{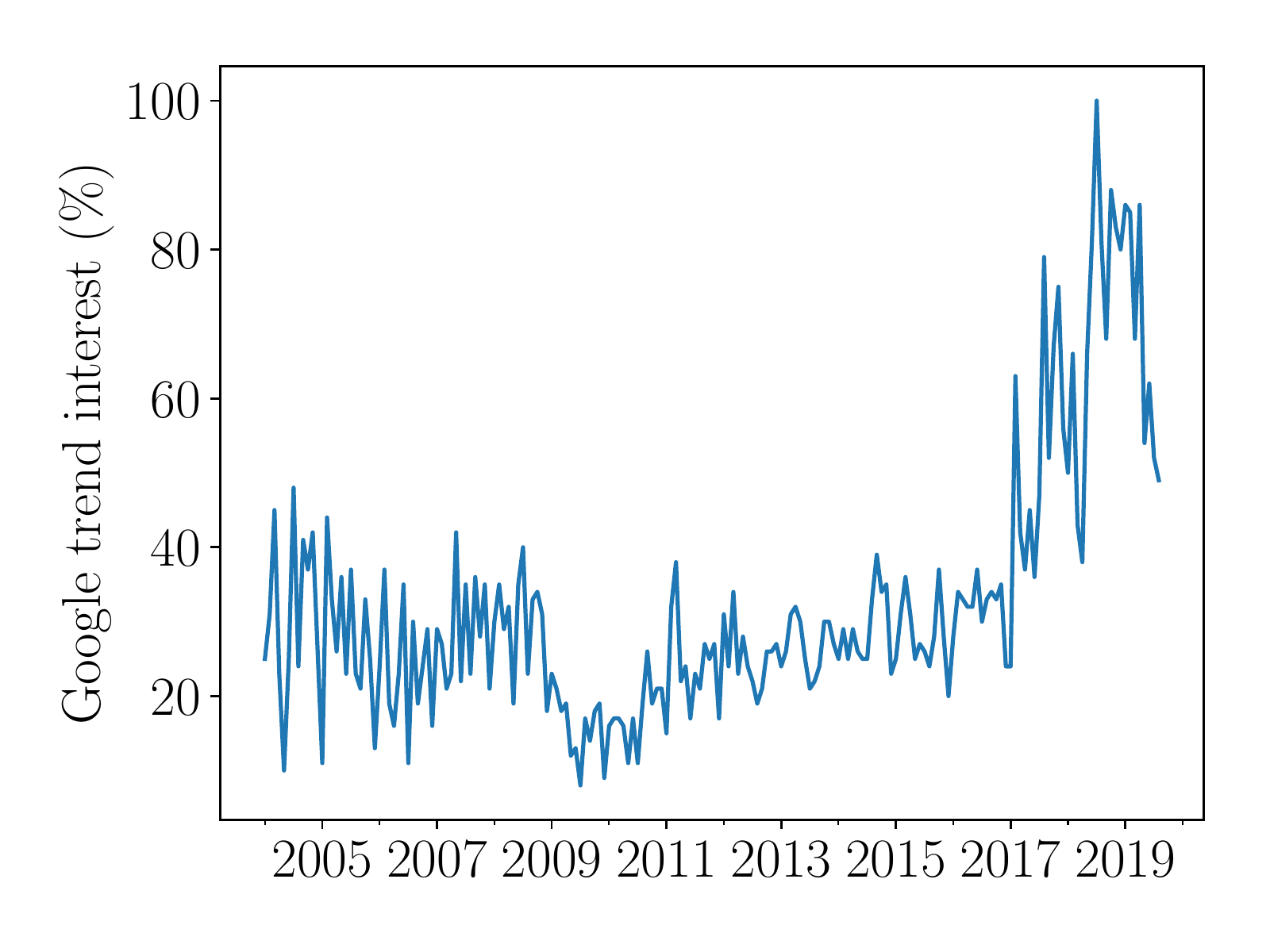}}
\par\end{centering}
\caption{Comparison of value of buybacks spent on S\&P 500 since 2004 and interest
measured by Google searches}

\label{fig:usinvestment}
\end{figure}

Figure \ref{fig:usinvestment} compares the amount spent on buybacks
by the S\&P 500 with a measure of interest in buybacks determined
through Google trends - a measure of the number of Google searches
worldwide.\footnote{Google trends interest data for the search term `share buyback' was
used from 2004-2019.} We see that there is a reasonable degree of correlation between the
two; furthermore, there is a peak in search activity in mid 2018.

\subsection{Quantifying share repurchases}

The following work aims to quantify explicitly how share repurchases
can be used to mechanically inflate (or deflate) a company's earnings
per share. A method is developed through an argument based on cash
flow analysis to examine how the buyback affects this, isolating a
few unitless quantities relating to both the current prevailing state
of the wider economy and a company's characteristics, all of which
can be obtained publicly, to quantify this effect. We primarily treat
the earnings as growing geometrically, though provide an alternative
analysis when the earnings are treated as growing arithmetically (providing
a more rigorous quantification of the uncertainty) in the Appendix. 

In addition we are able to explore the sensitivity of the growth in
earnings to the changes in the salient variables. Finally, we are
able use this analysis to estimate a company's natural growth rate
without this artificial inflation of the earnings. In particular,
we focus on the US market, examining the S\&P 500 and Apple as an
individual stock as case studies. 

\section{Accretive share repurchases}

We consider the effect on earnings as a consequence of share repurchases
at different prices. With similar expressions remarked in the literature
\cite{hribar2006stock,bens2003employee}, the following shows that
the repurchase is accretive (increases the earnings per share) only
if the price-earnings ratio, denoted $P_{E}$, of the traded company
is less than a certain value dependent on the interest and tax rate
($i$ and $t_{\text{Tax}}$ respectively). Or, put another way, if
the earnings yield (the reciprocal of the P/E ratio) is greater than
the after tax interest rate earned on the company's cash.

We have that the earning per share (for $N$ shares issued) for a
company can be expressed as follows 
\begin{align*}
E & :=\text{Earnings per share}\\
 & =\frac{(O+Ci)(1-t_{\text{Tax}})-m_{min}}{N}
\end{align*}
where $O$, $C$, and $m_{\text{min}}$ denote the trading profit,
cash reserves and minority charge respectively. If the company spends
$S$ to repurchase its shares at a price $P$, then with a dividend
of $d$ the earnings per share after the repurchase, $E'$, will be
adjusted as follows
\[
E'=\frac{(O+\left(C-S\left(1+\frac{d}{P}\right)\right)i)(1-t_{\text{Tax}})-m_{min}}{N-\frac{S}{P}}
\]
For the purposes of the following we assume that both the contribution
from dividends is negligible, i.e $d\ll P$, and that the minority
charge negligible, i.e. $m_{min}\approx0$. Finally, we let $\alpha=1-t_{\text{Tax}}$
denote the fraction of assets retained after tax. As a result, for
a repurchase to be accretive we require that $E'\ge E$ and so
\begin{align*}
(O+\left(C-S\right)i)\alpha N & \ge(O+Ci)\alpha\left(N-\frac{S}{P}\right)\\
(O+Ci)\alpha\frac{S}{P} & \ge Si\alpha N
\end{align*}
Which after rearrangement suggests that
\begin{align*}
\frac{(O+Ci)\alpha}{N}/P\ge i\alpha & \implies P_{E}=\frac{P}{E}\le\frac{1}{i\alpha}
\end{align*}
i.e. the price-earnings ratio is less than $\frac{1}{i\alpha}$. We
will, therefore, in the following refer to $P_{E}^{*}:=\frac{1}{\alpha i}$
as the `critical P-E ratio' and $P^{*}:=\frac{E}{\alpha i}$ as the
`critical price' given a company's earnings per share $E$. To ensure
proper units, we note that the period over which a company's earnings
per share is stated must correspond to the period over which interest
$i$ is paid. Unless otherwise stated, we assume that both time periods
are annualised. Nominal values for the critical PE ratio are shown
as a function of the interest rate $i\in[1\%,10\%]$ in Figure \ref{fig:peir}
and for value of tax rate between $10\%$ and $50\%$. We see that
for interest rates between 2-3\% with tax rates between 10-50\% then
the critical P/E is around 50.

Note that the long-term decline in interest rates over the last four
decades has had the effect of increasing the the critical P-E ratio
from approximately 15 in the late 1980's to the current level of roughly
50 in the late 2010s. A corollary of this effect as we shall come
on to quantify is that the boost to earnings per share for a given
proportion of shares repurchased has increased significantly over
the same period. 

\section{Earnings enhancement under buybacks}

We are interested in how the earnings per share after a buyback (or
sequence of buybacks) given the expenditure at time differs from that
without a buyback. In the following we will consider under a geometric
model for earnings growth what the natural earnings would look like
given knowledge of the size of the buyback and quantities reflecting
the state of the economy. For an interest rate of $i$ which we assume
is approximately constant, we define a company's earnings at time
$t$ as
\[
E_{t}=\alpha\frac{O_{t}+C_{t}i}{N_{t}}
\]

where $O_{t}$ is operating earnings and $C_{t}$ is the net interest
earning assets $\alpha=1-t_{\text{Tax}}$ which we assume is approximately
constant. Additionally, we assume that $O_{t}$ is the same in both
the presence and absence of a buyback. We model the interest earning
assets as 
\[
C_{T}=C_{t}I_{t,T}
\]
where $I_{t,T}$ is a continuously compounded interest rate process,
for example, $I_{t,T}=(1+\imath)^{T-t}$ for some constant $\imath\ge0$.
Throughout, $T$ will be used to refer to a terminal index and $t$
an initial point in time (sometimes 0).

We let $\gamma_{t}:=\frac{S_{t}}{P_{t}N_{t}}$ denote the fraction
paid of the market capitalisation in the buyback at time $t$ and
$m_{t}:=\frac{P_{t}}{P_{t}^{*}}$ as the definition of the share price
as a fraction of the critical price.
\begin{prop}
\label{prop:The-change-in}The difference in earnings at time $T$,
$E_{T}'$, relative to that without a buyback $E_{T}$ can be expressed
as 
\begin{equation}
E'_{T}=\frac{E_{T}-m_{t}\gamma_{t}I_{t,T}E_{t}}{1-\gamma_{t}}\label{eq:changeearningstt}
\end{equation}
 and we have the following approximation for the difference in earnings
with and without buybacks as 
\begin{align*}
E'_{T} & =E_{T}+\left(E_{T}-m_{t}I_{t,T}E_{t}\right)\sum_{n=1}^{\infty}\gamma_{t}^{n}\\
 & =E_{T}+\left(E_{T}-m_{t}I_{t,T}E_{t}\right)\gamma_{t}+\mathcal{O}(\gamma_{t}^{2})
\end{align*}

Proof in Appendix.
\end{prop}

We consider a simple example when the buyback is accretive so that
$m_{t}\in[0,1]$. and, additionally, under a simple growth strategy
where earnings at time $t$ are simply invested in a risk-free investment
we have that $E_{T}=I_{t,T}E_{t}$. In this case we have that 
\[
\frac{E_{T}'-E_{T}}{E_{T}}\ge\left(1-m_{t}\right)\gamma_{t}
\]
So we see that we may increase the earnings at a future time $T$
in the presence of a buyback at time $t$ by an amount at least proportional
to that spent on the buyback scheme $\gamma_{t}$. Such a relation
provides a significant incentive for managers with pay related to
earnings performance to increase the earnings per share through buying
back shares, up to the extent that they are accretive. We explore
this in more detail in the following.

\subsection{Instantaneous enhancement under buybacks}

\begin{figure}
\centerline{\subfloat[Critical PE ratio $i\alpha=i(1-t_{\text{Tax}})$ as a function of
interest rates.]{\includegraphics[scale=0.5]{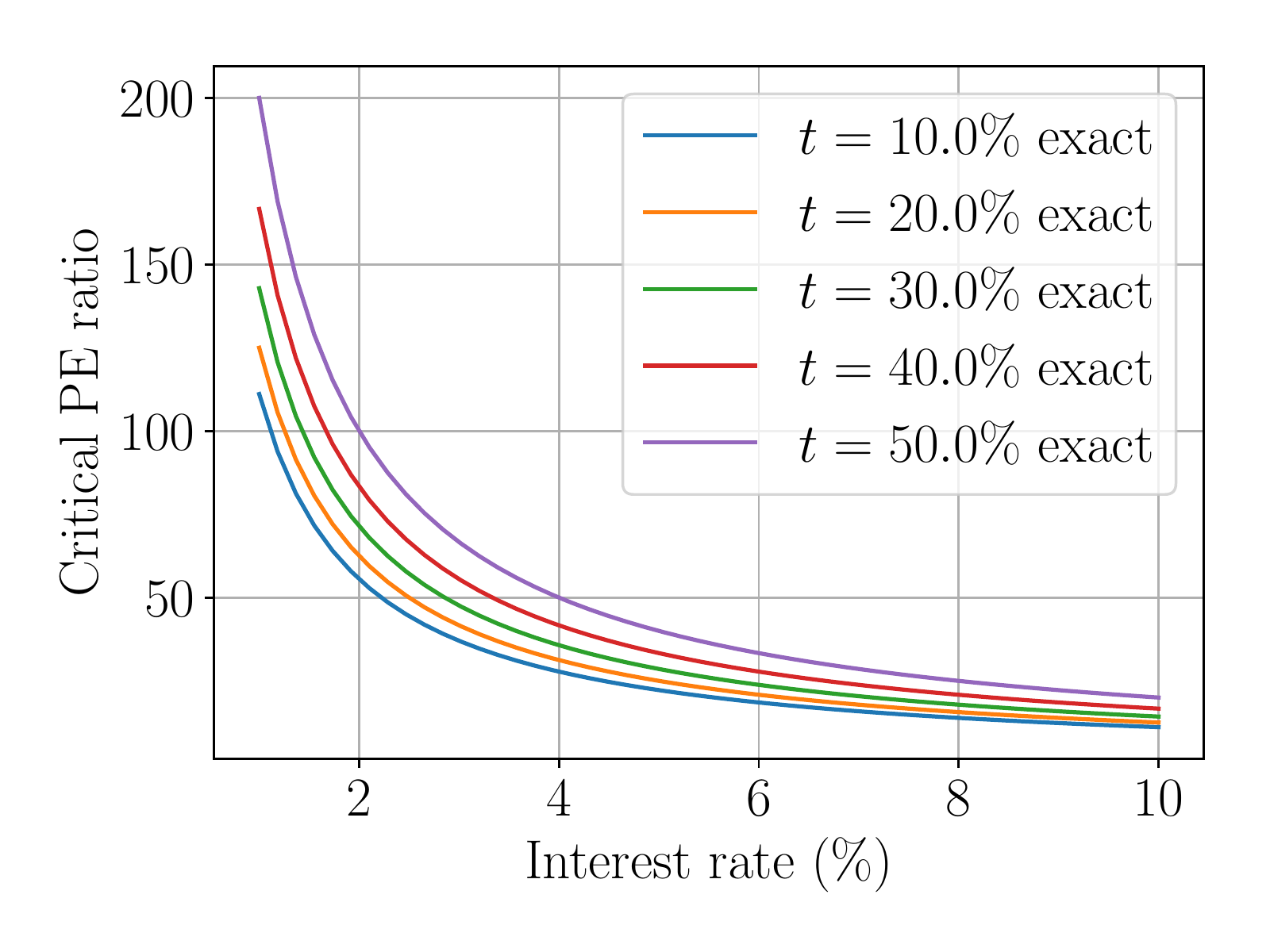}\label{fig:peir}}$\quad$\subfloat[Earnings enhancement $\Delta E$ in terms of price paid for repurchase.]{\includegraphics[scale=0.5]{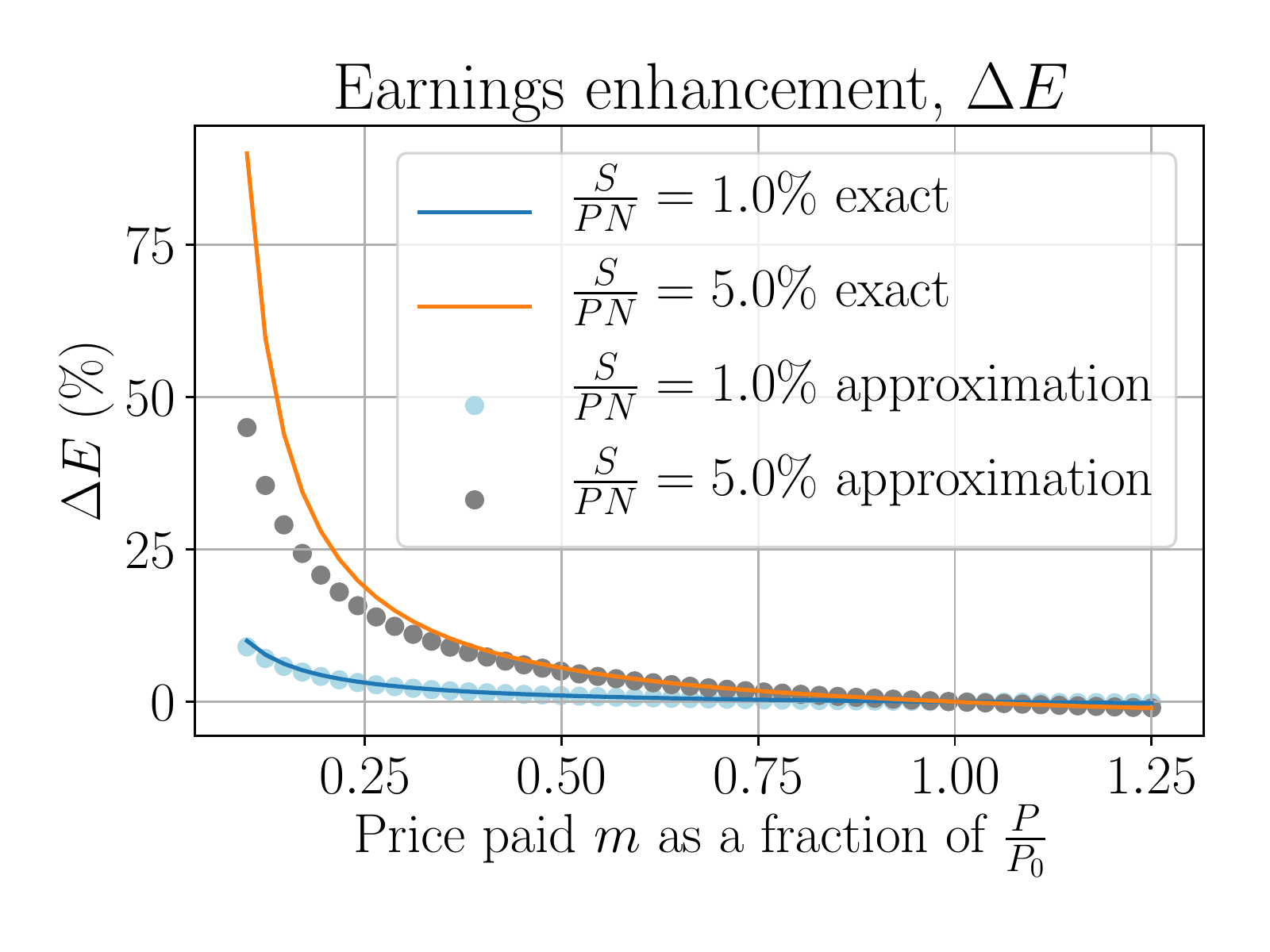}\label{fig:epsen}}}

\caption{Critical price and EPS enhancement}
\label{figepsenhfig1}
\end{figure}

We are able to consider the instantaneous change in earnings growth
as a result of the buyback through setting $T=t$. As a result we
see that by Proposition \ref{prop:The-change-in}, that the earnings
per share post-buyback can be expressed succinctly as
\[
E'=\frac{1-m\gamma}{1-\gamma}E
\]
dependency of $m=\frac{P}{P^{*}}$ and $\gamma$ on $t$ is suppressed
for notational convenience and we set $I_{t,t}=1$. Similarly, we
have that the relative change is then 
\begin{align}
\Delta E:=\frac{E'-E}{E} & =\frac{\gamma}{1-\gamma}(1-m)\label{eq:epsex}
\end{align}
We see as expected, for zero expenditure ($\gamma=0$) then the change
in earnings is 0. Furthermore, performing a series expansion of $\Delta E$
around $\gamma=0$, i.e. suggesting relative to the market capitalisation
the amount spent on the repurchase by the company is small we have
that up to an $o((m\gamma)^{2})$ term that
\begin{equation}
\Delta E\approx\left(\frac{1}{m}-1\right)m\gamma=\frac{S}{P^{*}N}\left(\frac{1}{m}-1\right)\label{eq:epsapprox}
\end{equation}
In particular, if we further assume that the buyback is accretive,
i.e. $P_{E}\le P_{E}^{*}$, then we have that both $m,\gamma\in[0,1]$
and that the remainders in the series expansion are positive, suggesting
that in fact $\Delta E\ge\frac{S}{P^{*}N}\left(\frac{1}{m}-1\right)$
so that the approximation is conservative in this sense. This linearisation
shows how the relative change in earnings is affected as a function
of the proportion spent on the critical price. The gains (or losses)
of earnings as a result of the buyback under this normalisation are
plotted in Figure \ref{figepsenhfig1} as a function of the price
paid for the shares for both the exact expression \ref{eq:epsex}
and the first order approximation \ref{eq:epsapprox}. As can be seen,
the linearisation is considerably accurate for $\frac{S}{P^{*}N}\approx1\%$
and for a large range of $m$. We also see that the improvement to
earnings rises quite steeply as $m$ approaches 0, demonstrative of
the highly nonlinear relationship between the two. 

\subsection{Earnings enhancement under geometric growth in earnings}

We are now interested in the value of earnings at some future time
$T$ given a buyback for $\gamma_{t}$ at time $t<T$. In the following,
we let $E_{t}$ denote the \emph{natural earnings} under the absence
of a buyback scheme and define the normalised natural earnings $\hat{x}$
from time $t\ge0$ to $T\ge t$ implicitly as satisfying
\[
E_{T}=\hat{x}_{t,T}E_{t}
\]
In this case we have that from Proposition \ref{prop:The-change-in}
that

\begin{align*}
E_{T}' & =\frac{E_{T}-m_{t}\gamma_{t}I_{t,T}E_{t}}{1-\gamma_{t}}\\
 & =\frac{1}{1-\gamma_{t}}\left(E_{T}-I_{t,T}\frac{E_{t}}{P_{t}^{*}}\frac{S_{t}}{N_{t}}\right)\\
 & =\frac{1}{1-\gamma_{t}}\left(1-\frac{I_{t,T}}{\hat{x}_{t,T}}\frac{S_{t}}{P_{t}^{*}N_{t}}\right)E_{T}
\end{align*}

Finally, substituting again $\gamma_{t}=\frac{S_{t}}{P_{t}N_{t}}$
as the proportion spent on share buybacks relative to the market capitalisation
and substitute $P_{t}=mP_{t}^{*}$ we can express this neatly as 
\begin{equation}
E_{T}'=\frac{1}{1-\gamma_{t}}\left(1-m\gamma_{t}\frac{I_{t,T}}{\hat{x}_{t,T}}\right)E_{T}\label{eq:epsenhquant}
\end{equation}
We consider the special case of $t=0$ and simple constant geometric
growth models both for interest rates and growth, i.e. that $\hat{x}_{0,T}=(1+\xi)^{T}$
and $I_{0,T}=(1+\imath)^{T}$ for two constants $\xi,\imath\ge0$.
In this case letting $\gamma$ denote the normalised value of the
buyback at $t=0$ then

\[
E_{T}'=\frac{1}{1-\gamma}\left(1-m\gamma\left(\frac{1+\imath}{1+\xi}\right)^{T}\right)E_{T}
\]

As we can see, if we consider the `immediate' change in EPS by setting
$T=0$, we recover the earnings enhancement as detailed above. Such
a formula reveals a number of important characteristics of share buybacks.
Firstly, to a degree as expected, as the interest rate $\imath$ increases
the future enhancement diminishes, suggesting that the company would
have been better retaining the cash rather than repurchasing shares,
as measured by the earnings per share. Furthermore, we see that up
to an $o(\gamma^{2})$ term then a series expansion of $E'_{T}$ around
$\gamma=0$ provides the approximation
\begin{equation}
\frac{E_{T}'-E_{T}}{E_{T}}\approx\left(1-m\left(\frac{1+\imath}{1+\xi}\right)^{T}\right)\gamma.\label{eq:taylor future}
\end{equation}

\begin{figure}
\begin{centering}
\includegraphics[scale=0.5]{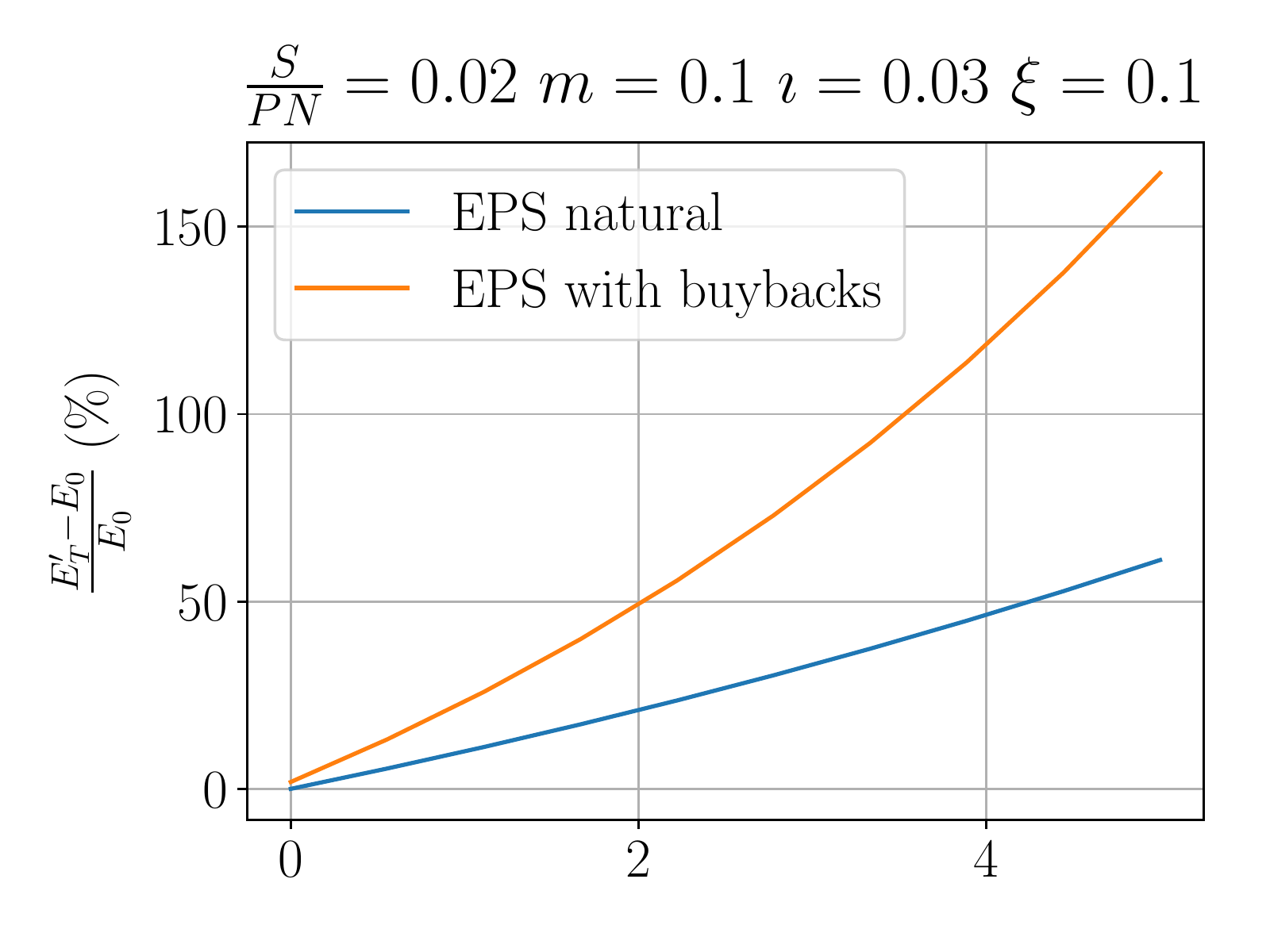}
\par\end{centering}
\caption{Example earnings enhancement over a period of 5 years, normalised
by the initial earnings $E_{0}$.}
\label{fig:epsenthroughtime}
\end{figure}

We refer to the boost arising from buybacks to the earnings as `the
earnings enhancement'. The series approximation of (\ref{eq:taylor future})
is conservative in that the remainder is positive for $\gamma>0$.
We explore nominal values of $\gamma$ in the next section. We note
that provided $\xi>\imath$ the earnings enhancement converges exponentially
towards $\lim_{T\rightarrow\infty}\left(1-m\left(\frac{1+\imath}{1+\xi}\right)^{T}\right)\gamma=\gamma$
through time suggesting that despite the initial cost in repurchasing
the shares, the effect on the earnings can diminish relatively quickly
through time.

We plot the EPS enhancement through time for some representative values
of $m$, $x$ and $i$. Based on the current Federal Funds rate of
$2\%$ we use this value for $i$. We then consider $x=0.1$ with
$m=0.1$, i.e. an annual growth rate of $10\%$ and shares repurchased
at $10\%$ of the critical price and $\frac{S}{P^{*}N}=2\%$. As will
be seen, such values are representative of the buyback scheme undertaken
by Apple since 2011/2012. For these two values of $m$ we compare
$\frac{E_{T}'-E_{0}}{E_{0}}$ with $\frac{E_{T}-E_{0}}{E_{0}}$, plotting
the results in \ref{fig:epsenthroughtime}.

\subsection{Earnings enhancement under buyback program}

We here illustrate that under benign conditions (i.e. accretive buybacks)
the growth in earnings increases geometrically \emph{on top of} a
geometric earnings growth as per the special case considered previously. 

Consider the affect on earnings per share after a series of buybacks
occurring at regular intervals. We have that at time $t$ after a
buyback at time 0 that the earnings enhancement above the natural
EPS at time $t$ can be expressed equivalently as
\begin{align*}
\Delta E_{t} & =\frac{\gamma}{1-\gamma}\left(1-m\left(\frac{1+\imath}{1+\xi}\right)^{t}\right)
\end{align*}

We consider a buyback program with repurchases occurring at intervals
$t=cn$ for $n=1,2,...,T$ and $c>0$. In this case we see that
\begin{align}
E_{n}' & =\prod_{k=0}^{n}(1+\Delta E(c(n-k)))E_{n}\nonumber \\
 & =\prod_{k=0}^{n}\left(\frac{1}{1-\gamma}\left(1-m\gamma\left(\frac{1+\imath}{1+\xi}\right)^{c(n-k)}\right)\right)(1+\xi)^{cn}E_{0}\nonumber \\
 & =\left(\frac{(1+\xi)^{c}}{1-\gamma}\right)^{n}\prod_{k=0}^{n}\left(1-m\gamma\left(\frac{1+\imath}{1+\xi}\right)^{c(n-k)}\right)E_{0}\nonumber \\
 & =\left(\frac{(1+\xi)^{c}}{1-\gamma}\right)^{n}\prod_{k=0}^{n}\left(1-m\gamma\left(\frac{1+\imath}{1+\xi}\right)^{ck}\right)E_{0}\label{eq:buybackt}
\end{align}
for low interest rates, i.e. $\xi\ge\imath$ we see that as $m,\gamma\in[0,1]$
then $\left(1-m\gamma\left(\frac{1+\imath}{1+\xi}\right)^{ck}\right)\ge\left(1-m\gamma\left(\frac{1+\imath}{1+\xi}\right)^{c}\right)$
as $k\ge1$. We see that 
\begin{align*}
\frac{E_{n}'}{E_{0}} & \ge\left(\frac{(1+\xi)^{c}}{1-\gamma}\right)^{n}\left(1-m\gamma\left(\frac{1+\imath}{1+\xi}\right)^{c}\right)^{n}\\
 & \ge(1+\xi)^{cn}\left(1+d\gamma\right)^{n}
\end{align*}
where the inequality arises from a series approximation up to an $o(\gamma^{2})$
term, as before, with the constant $d$ independent of $\gamma$.
In periods of low interest rate relative to company growth the increase
in earnings per share resulting from a buyback scheme is at least
geometrically increasing both with the growth of the company $\xi$
and with the proportion spent on buybacks $\gamma$. We examine the
behaviour in the limit as $n\rightarrow\infty$ in which case we see
that 
\begin{align*}
\lim_{n\rightarrow\infty}\frac{1}{n}\log\left(\frac{E_{n}'}{E_{0}}\right) & =\log\left(\frac{(1+\xi)^{c}}{1-\gamma}\right)+\lim_{n\rightarrow\infty}\frac{1}{n}\sum_{k=0}^{n}\log\left(1-m\gamma\left(\frac{1+\imath}{1+\xi}\right)^{ck}\right)\\
 & =\log\left(\frac{(1+\xi)^{c}}{1-\gamma}\right)
\end{align*}
in contrast, the simple model without buybacks, in which case $\frac{1}{n}\log\left(\frac{E_{n}}{E_{0}}\right)=(1+\xi)^{c}$,
and that asymptotically at least, that the annual growth artificially
inflates by approximately $(1/(1-\gamma))^{n}\approx(1+\gamma)^{n}$
purely as a result of buybacks. %

\begin{figure}
\centerline{\subfloat[Earnings per share in the presence of buybacks $\frac{E_{n}'}{E_{0}}$
(solid lines) and natural earnings per share $\frac{E_{n}}{E_{0}}$
(dashed lines) for a range of values of the natural growth rate $\xi$.]{\includegraphics[scale=0.5]{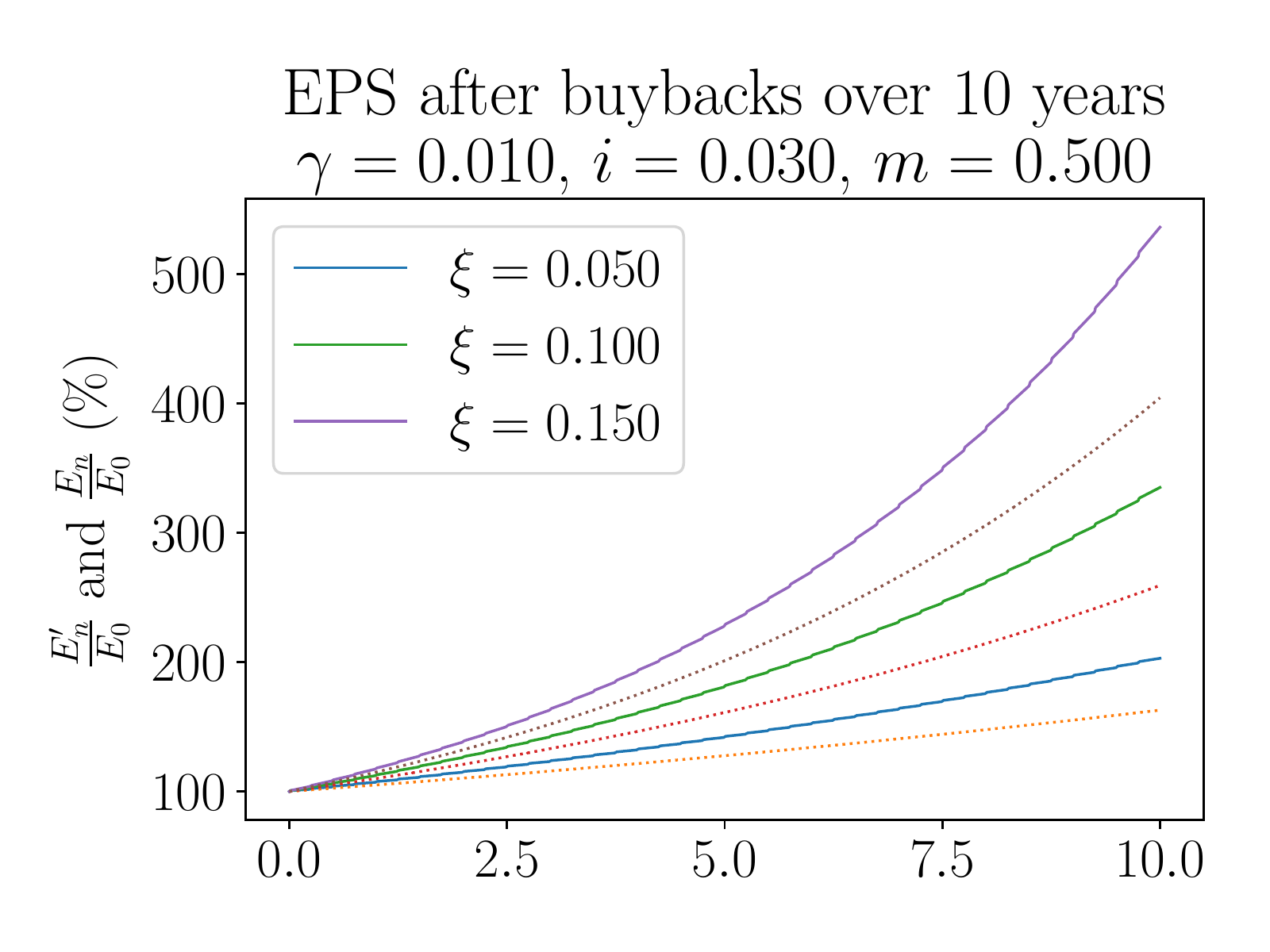}\label{fig:cheapcompoundbb}}$\quad$\subfloat[Difference between $\frac{E'_{n}}{E_{0}}$ and $\frac{E_{n}}{E_{0}}$
is percentage growth, for each of the values of $\xi_{n}$ in Figure
\ref{fig:cheapcompoundbb}.]{\includegraphics[scale=0.5]{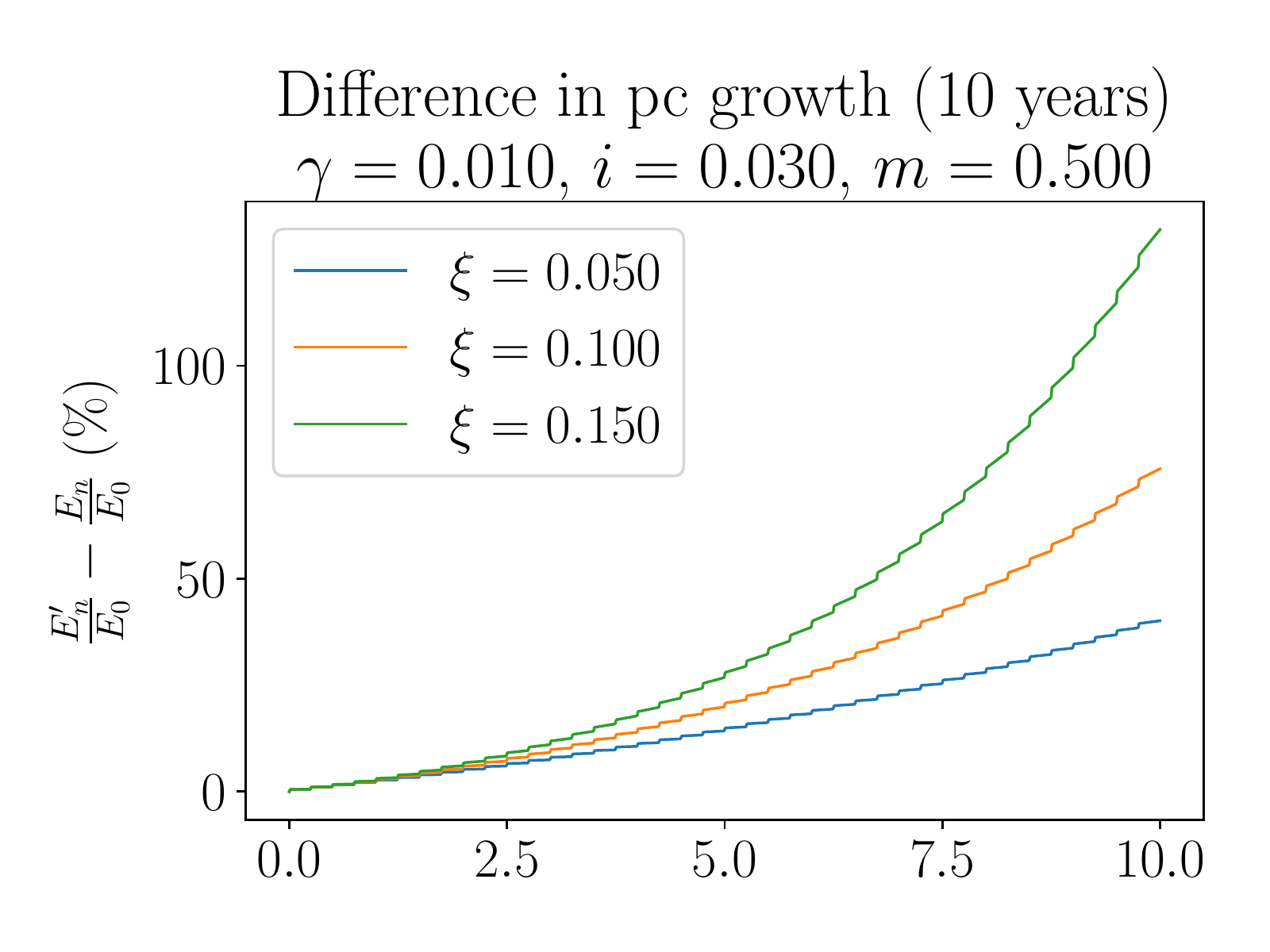}\label{fig:mediumbuybacl}}}

\caption{Compounded EPS after a sequence of accretive buybacks}
\end{figure}

Finally we plot how the sequence of buybacks affects the EPS for some
toy-values. We consider firstly $m=0.5$, i.e. where the company's
$P_{E}$ ratio shares is $50\%$ of $P_{E}^{*}$ the critical value,
in Figure \ref{fig:cheapcompoundbb}. It can be seen after a sequence
of buybacks occurring quarterly, we have that the EPS is between two
and three times larger compared to the vanilla growth in EPS, as provided
by $E_{n}=(1+\xi)^{n}E_{0}$. 

\section{Methodology}

In the following, we focus on the geometric model for earnings under
the effect of a buyback program on real data, examining primarily
to what extent accretive buybacks increase earnings per share. Of
central interest is what are representative values of $m$ and $\gamma$
in a real setting for which we appeal to the US stock market. We focus
on the situation of a sequence of regular buybacks, with the normalised
growth in earnings given by (\ref{eq:buybackt}) 
\begin{equation}
E_{n}'=\left(\frac{(1+\xi)^{c}}{1-\gamma}\right)^{n}\prod_{k=0}^{n}\left(1-m\gamma\left(\frac{1+\imath}{1+\xi}\right)^{ck}\right)E_{0}\label{eq:latent}
\end{equation}
Estimates of $m=\frac{P}{P^{*}}=\frac{P_{E}}{P_{E}^{*}}$ rely on
values of the critical $P_{E}^{*}=\frac{1}{i(1-t_{Tax})}$, which
in the case of US stocks we take the 10-year Treasury Constant Maturity
bond yield and consider both the prescribed US corporate tax rate
and the effective tax rate. 

We will use nonlinear least squares to estimate $\xi$ - the natural
growth in earnings - from the data. In a statistical setting, the
optimisation can be thought of as fitting normally distributed observations,
with mean given by \ref{eq:latent} and constant variance. We therefore
minimise 
\[
(\xi',E_{0}):=\arg\min_{(\xi,E_{0})}\sum_{n=0}^{n'}\left(R_{n}-E_{n}'\right)^{2}
\]
where $R_{n}$ is the realised earnings and $E_{n}'$ is those under
the model in Equation \ref{eq:latent}. 

\begin{figure}
\centerline{\subfloat[US interest rate and tax rates. (- -) Corporate tax rate (-) Effective
tax rate]{\includegraphics[scale=0.5]{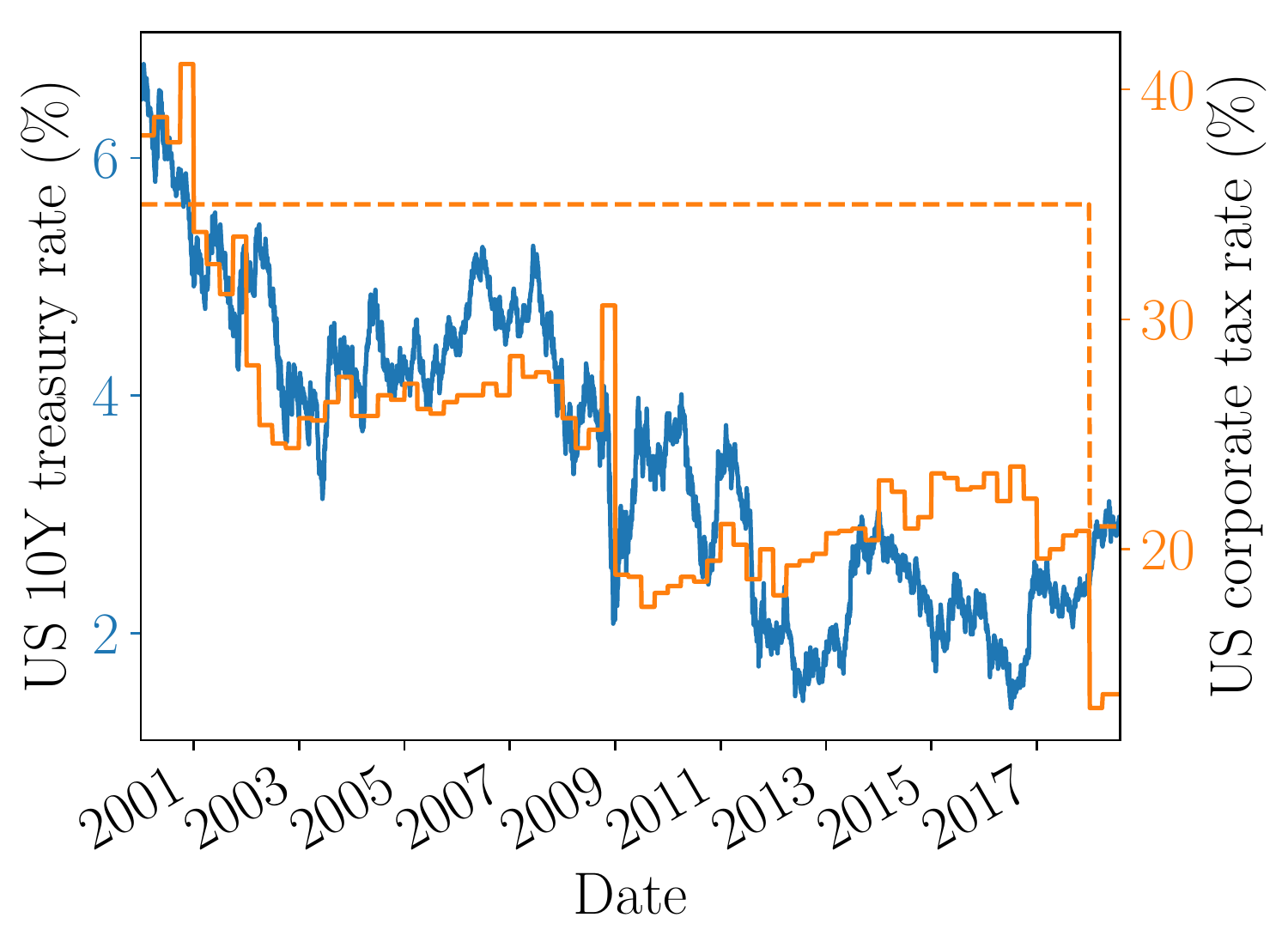}\label{fig:irtax}}$\quad$\subfloat[US critical $P_{E}$ with the two tax rates considered]{\includegraphics[scale=0.5]{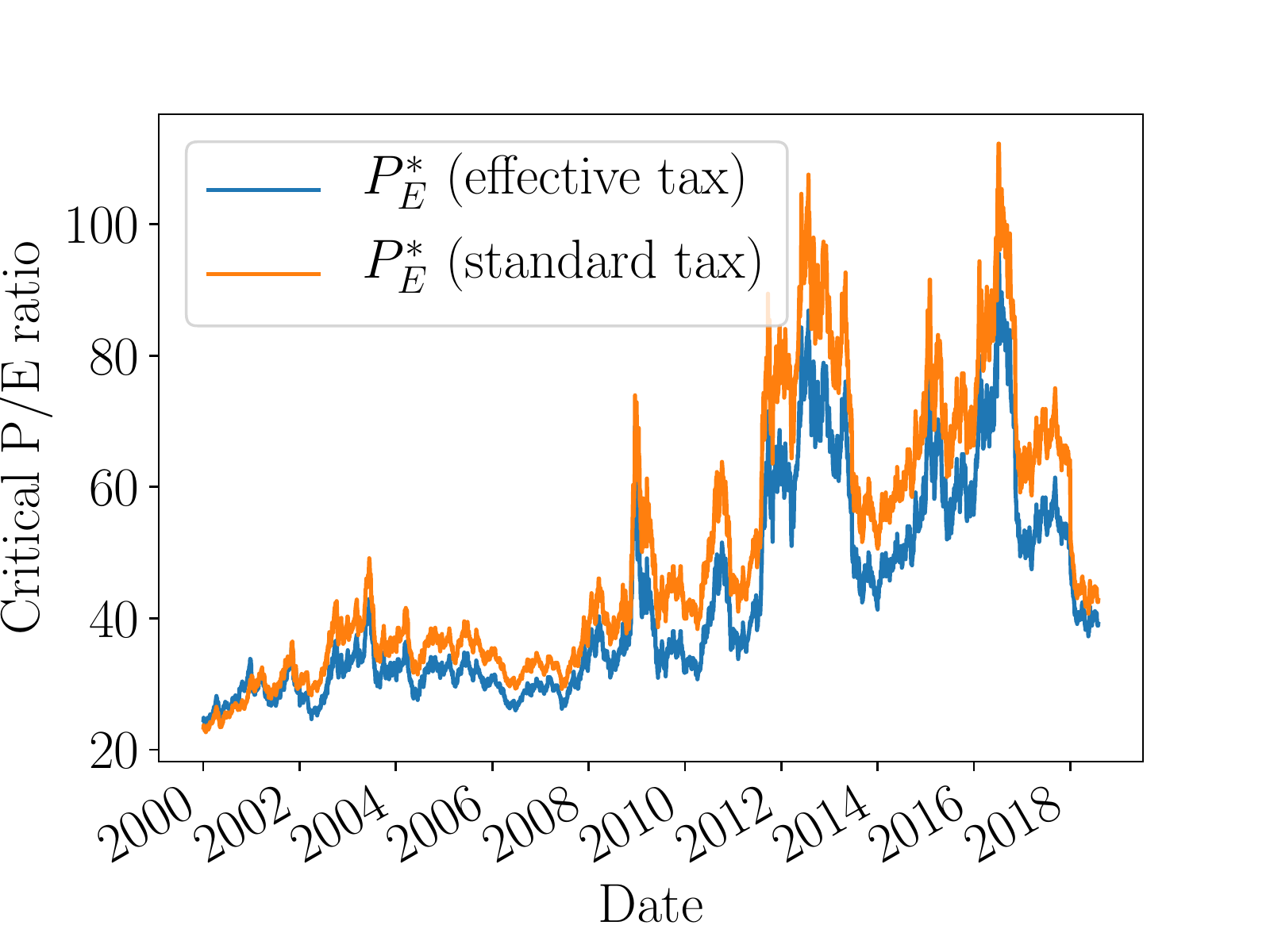}\label{fig:uscrit}}}

\caption{Critical P/E, $P_{E}^{*}$, for US market. Interest rate and tax rate
data taken from the Federal Reserve. }
\label{fig:criticalpe}
\end{figure}

Following methodology similar to \cite{econ2018yardeni} we estimate
the effective tax rate using the difference between pre-tax and post-tax
profits normalised by the pre-tax profits, with the results shown
in Figure \ref{fig:criticalpe}. 

Figure \ref{fig:irtax} shows both the 10Y bond yield and the two
tax rates. Excluding the recent tax reforms we see that the corporate
tax has been approximately constant over this period at approximately
40\%, whereas the effective tax rate is significantly lower. Importantly,
this has an effect on the critical P/E in that by definition it is
monotonically decreasing in the tax rate. As such, we plot the critical
P/E using both these values in Figure \ref{fig:uscrit} where we see
that while there is a discrepancy between the two tax rates, the discrepancy
is relatively mild. More strikingly, from Figure \ref{fig:uscrit}
we see an overall positive trend in critical P/E over this period,
suggesting that for low-growth companies, in terms of relatively static
$P_{E}$ ratios, the result of a buyback is more likely to be accretive
and the earnings per share are more likely to be increased after a
buyback. %

\subsection{Apple}

\begin{figure}
\centerline{\subfloat[Estimates of $\gamma$ using quarterly buyback data, (- -) median
($\cdot\cdot$) lower and upper 20\%-iles.]{\centering{}\includegraphics[scale=0.5]{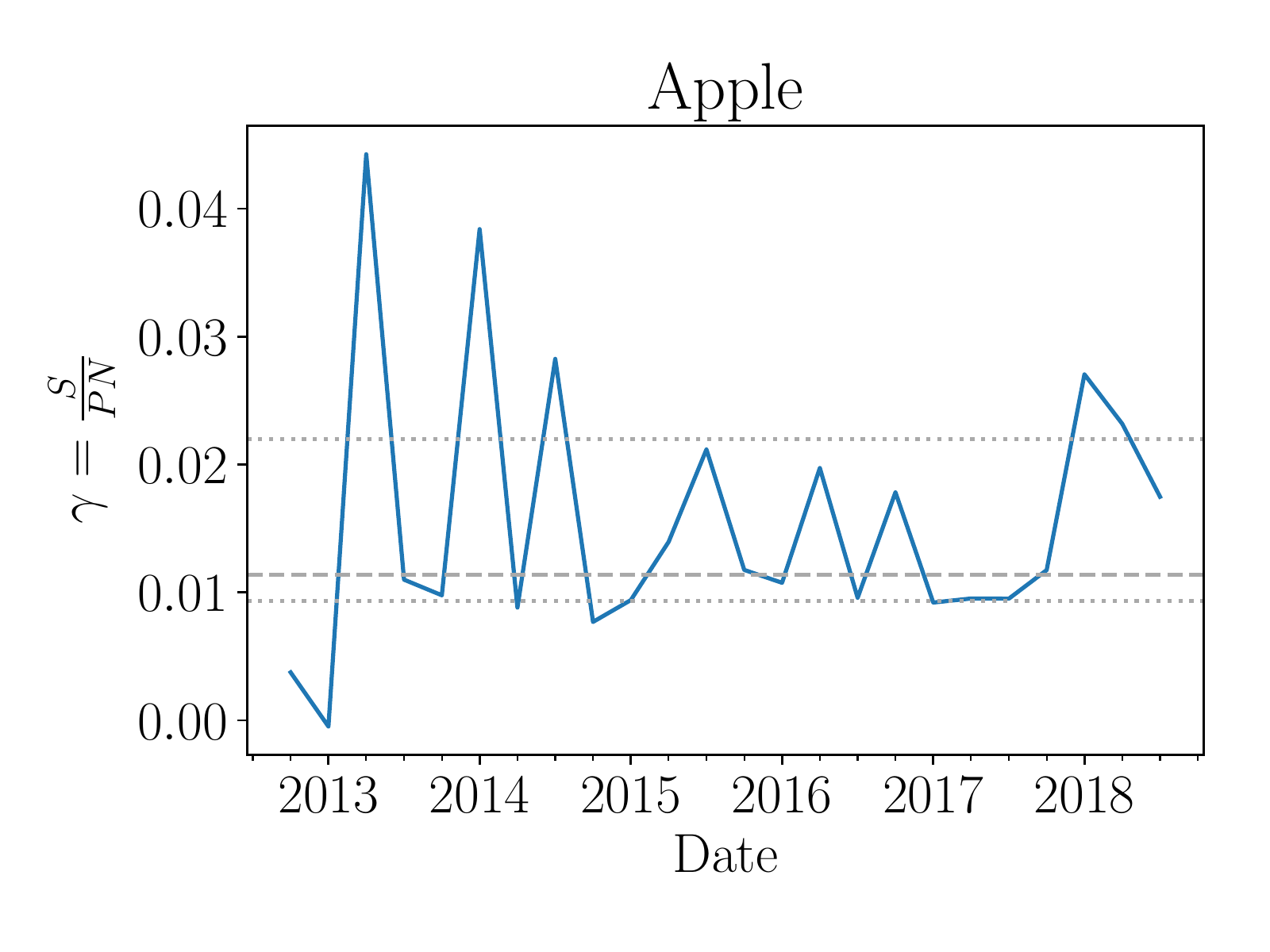}}\subfloat[Estimate of ratio of critical $P/E$ using the effective tax rate
to calculate the critical $P_{E}^{*}$ (- -) median ($\cdot\cdot$)
lower and upper 20\%-iles.]{\centering{}\includegraphics[scale=0.5]{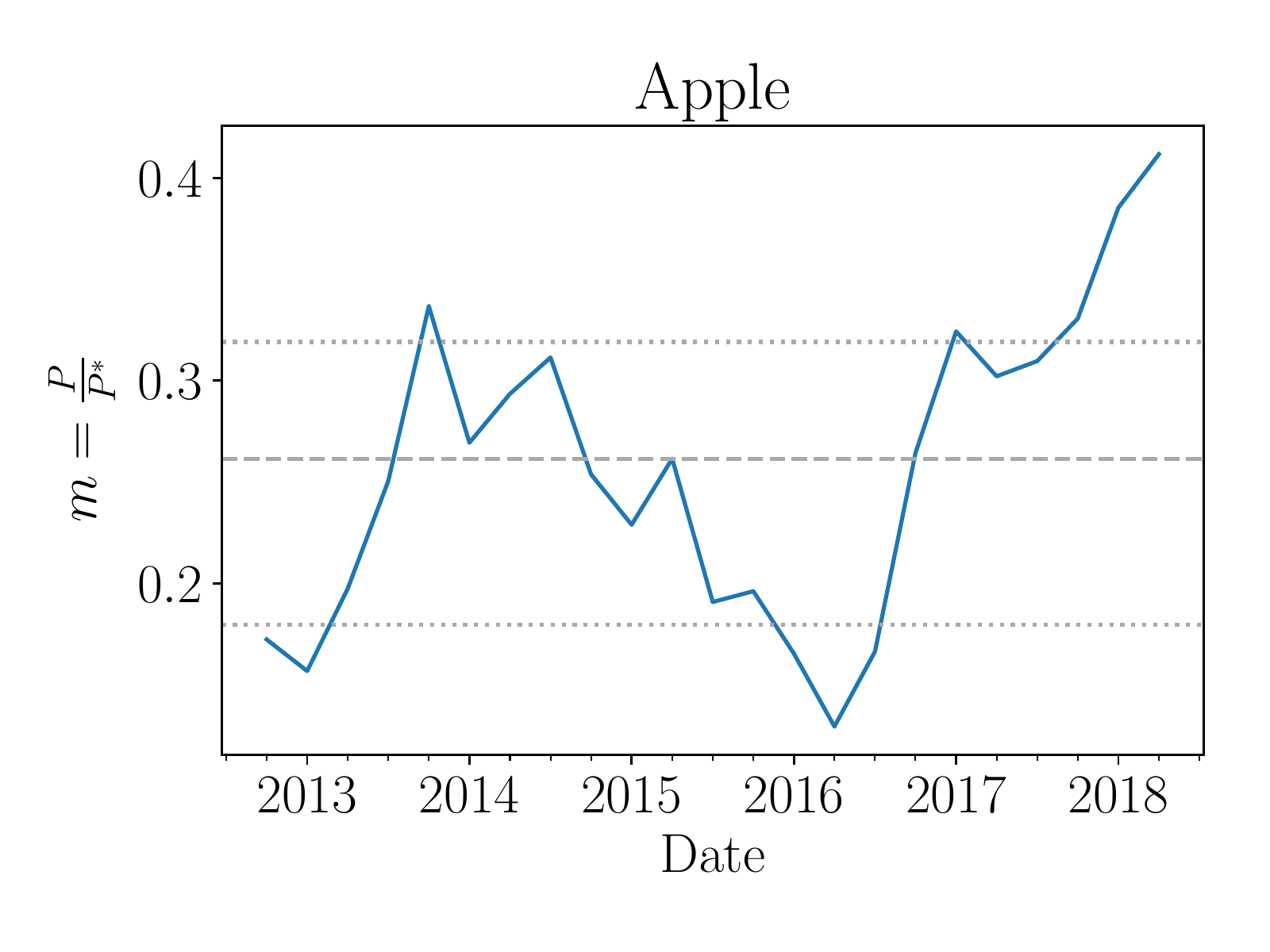}}}

\caption{Apple data used to estimate $m$ and $\gamma$}
\label{fig:applegam}
\end{figure}

\begin{figure}
\centerline{\subfloat[Value of buybacks]{\centering{}\includegraphics[scale=0.5]{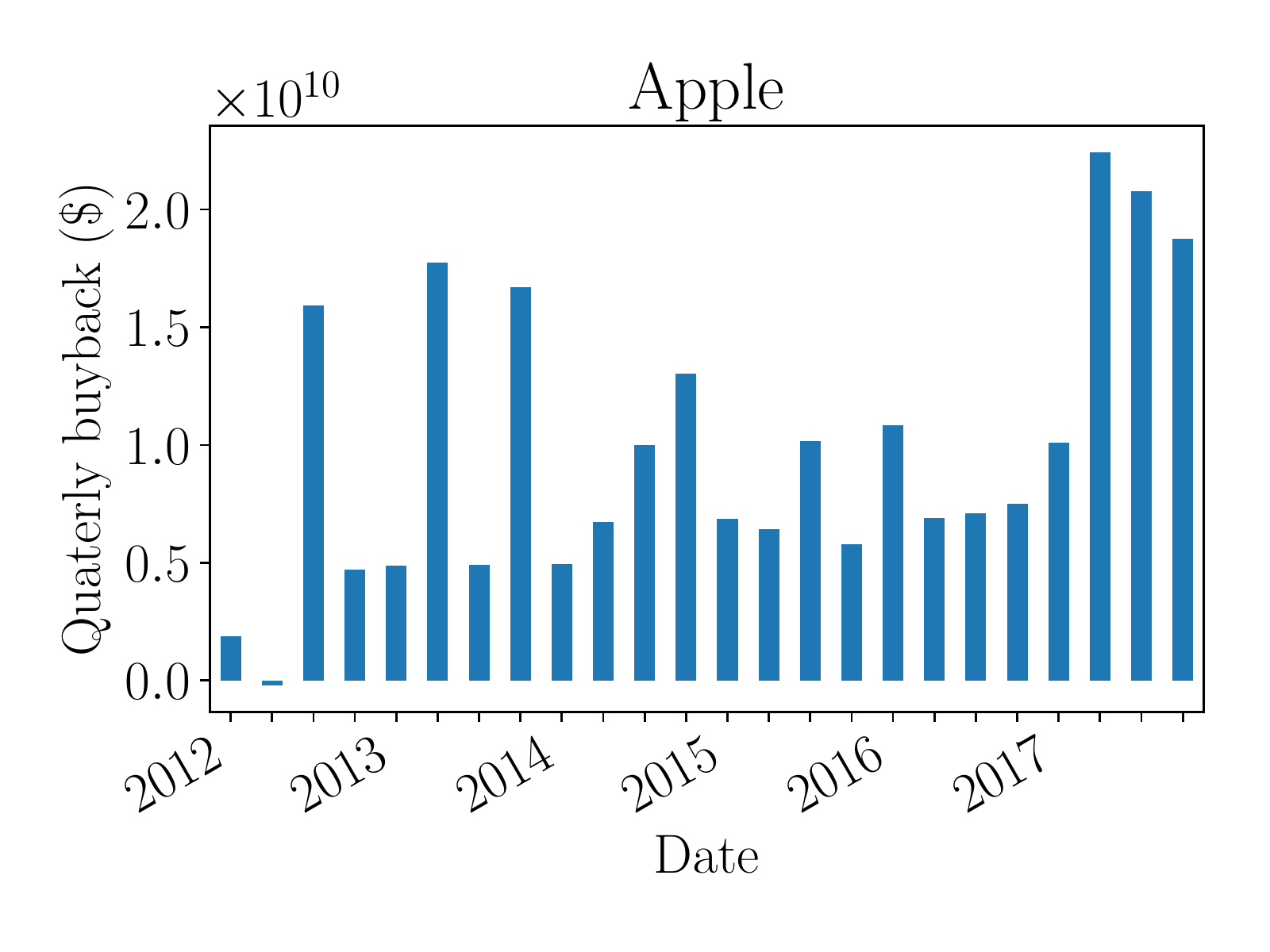}}\subfloat[Comparison between realised earnings per share $R_{n}$ used to fit
the model and earnings per share in the presence of buybacks $E_{n}'$
and the natural earnings per share $E_{n}=(1+\xi)^{n}$.]{\centering{}\includegraphics[scale=0.5]{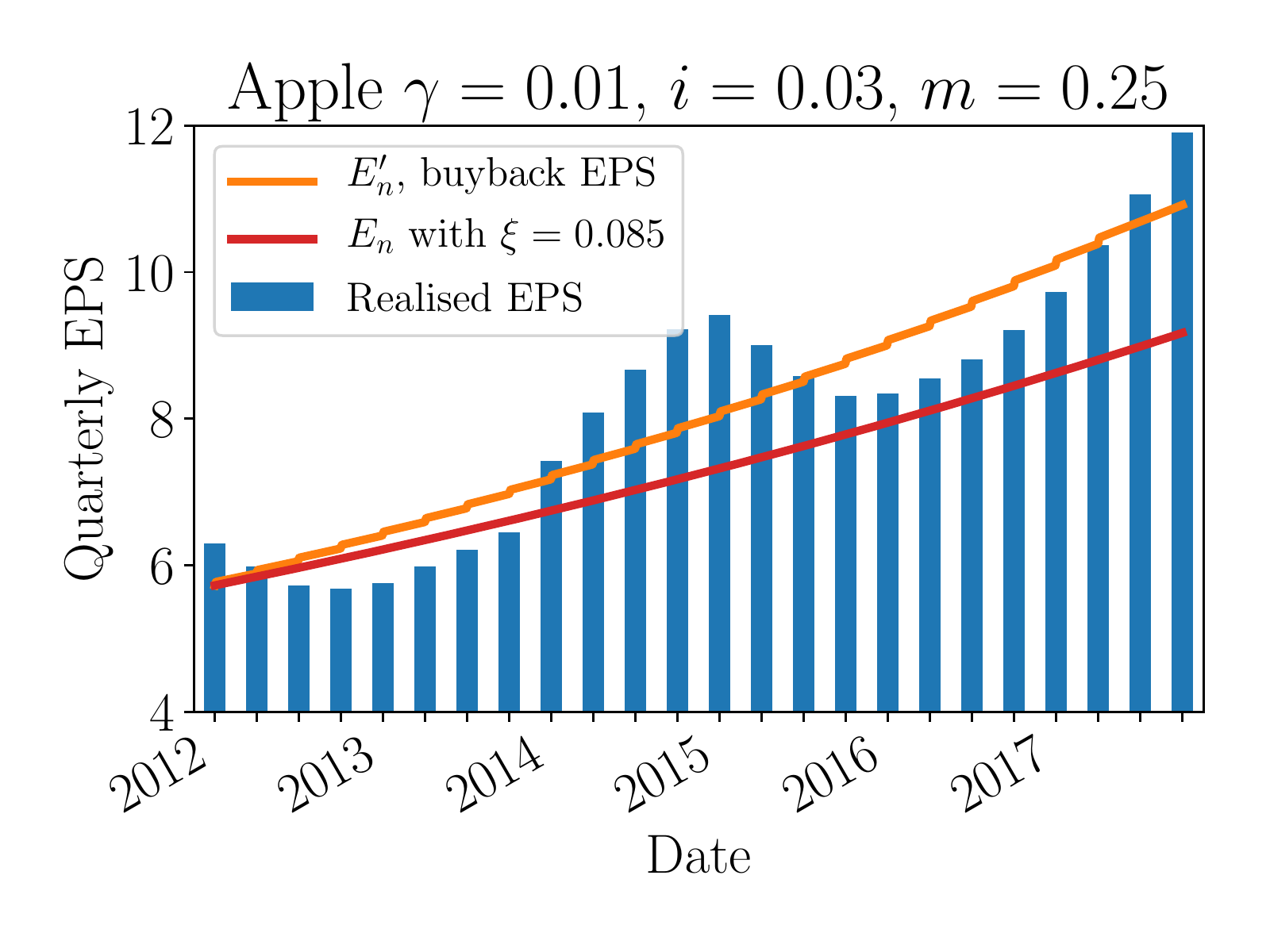}\label{fig:appleepsgrowth-1-1}}}

\caption{Comparison of realised earnings growth and natural growth in the absence
of buybacks for Apple shares over a recent 6 year window.}
\label{fig:applegam-1}
\end{figure}

We focus on Apple during the period of buybacks initiating in 2013,
until the present day, comparing the natural and observed growth rates
through the relation of Equation \ref{eq:buybackt} after a sequence
of buybacks occurring quarterly at regular intervals. The free parameters
in the model are $m=\frac{P}{P^{*}}$, the fraction of the critical
price paid, $\gamma=\frac{S}{PN}$ the fraction of the market capitalisation
spent on buybacks, and $x$ the expected natural growth rate in EPS.
We estimate both $m$ and $\gamma$ using the historical data, shown
in Figure \ref{fig:applegam}, plotting the relevant quantities, value
of the buyback (paid quarterly), with $m$ shown to be in the range
0.2 to 0.4 and $\gamma$ shown to vary between 0 and 0.04, with a
mean of $0.01$. Taking $m$ and $\gamma$ as the central values of
$0.25$ and $0.01$ respectively we compare the historical growth
in earnings (shown in $\%$ increase since the start of the period
- $1^{st}$ January 2012) for a natural growth rate of $x=0.08$.
The results are shown in Figure \ref{fig:appleepsgrowth-1-1}, where
we see that over this 5-year period the observed growth in earnings
is approximately 188\%, however, with a natural estimated growth of
only 164\%, suggesting that as much of a quarter of the earnings growth
over the period ($1-\frac{64}{88}=27\%$ 2s.f.) can be attributable
to share repurchases.

\subsection{S\&P 500 }

\begin{figure}
\centerline{\subfloat[$m=\frac{P}{P^{*}}$ for S\&P 500]{\centering{}\includegraphics[scale=0.5]{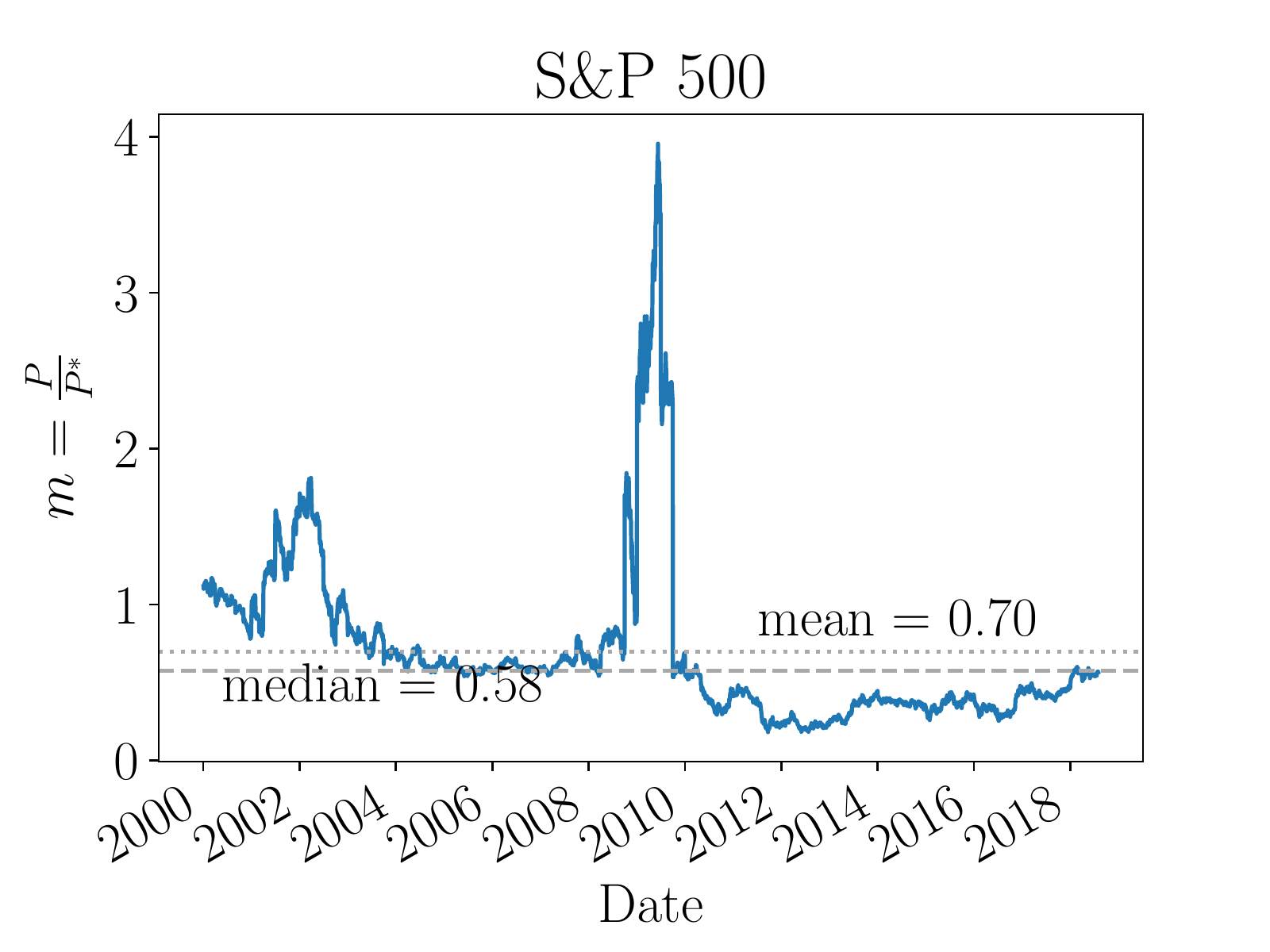}\label{fig:spm}}\subfloat[Proportion spent on buybacks of market capitalisation, $\gamma=\frac{S}{NP}$,
for S\&P 500]{\centering{}\includegraphics[scale=0.5]{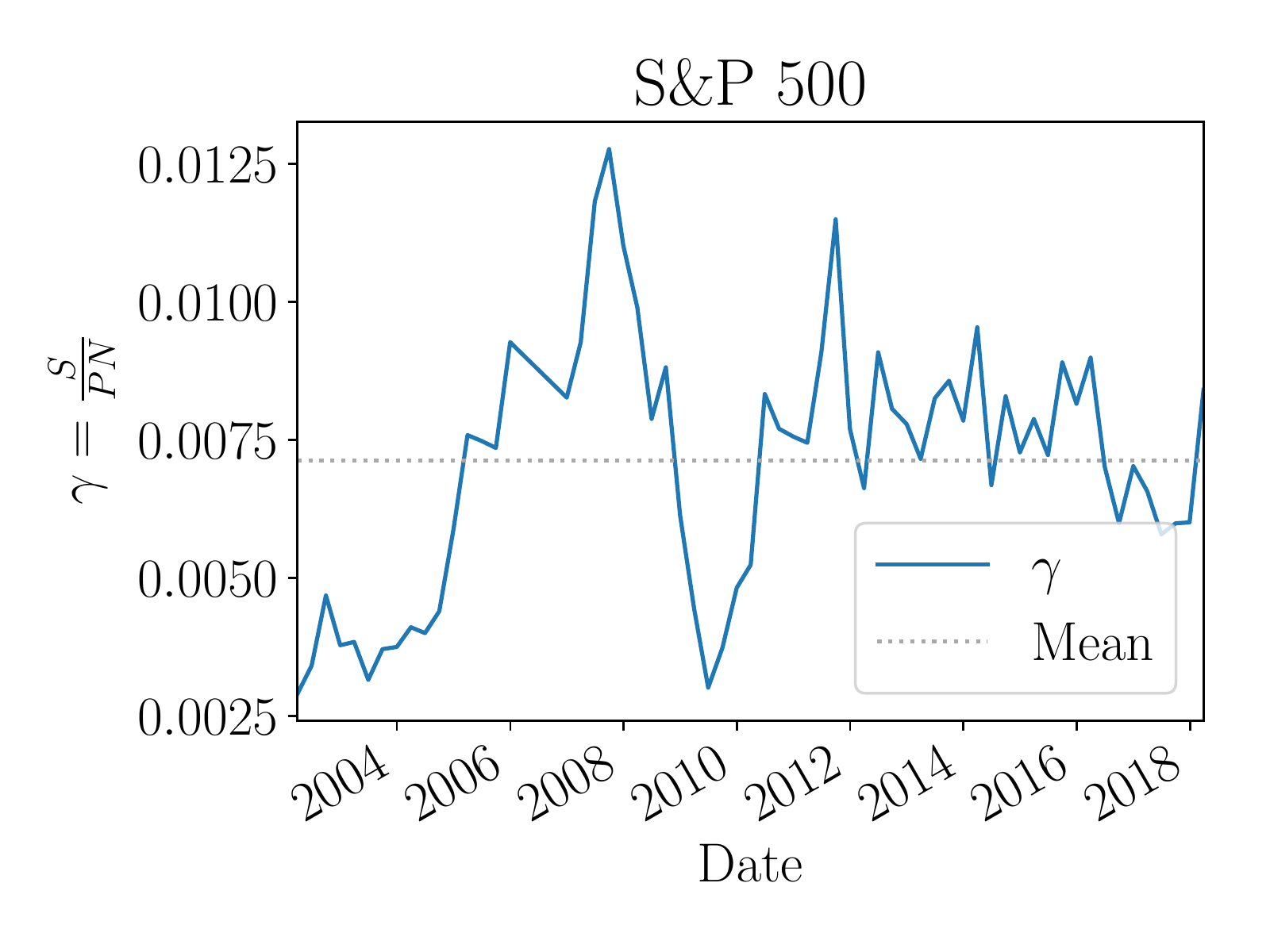}\label{fig:spgamma}}}

\caption{Values of $m$ and $\gamma$ for S\&P 500 from the quarter ending
in $31^{st}$ March 2002 to the quarter ending $29^{th}$ March 2018.}
\label{fig:applegam-1-1}
\end{figure}

\begin{figure}
\begin{centering}
\includegraphics[scale=0.7]{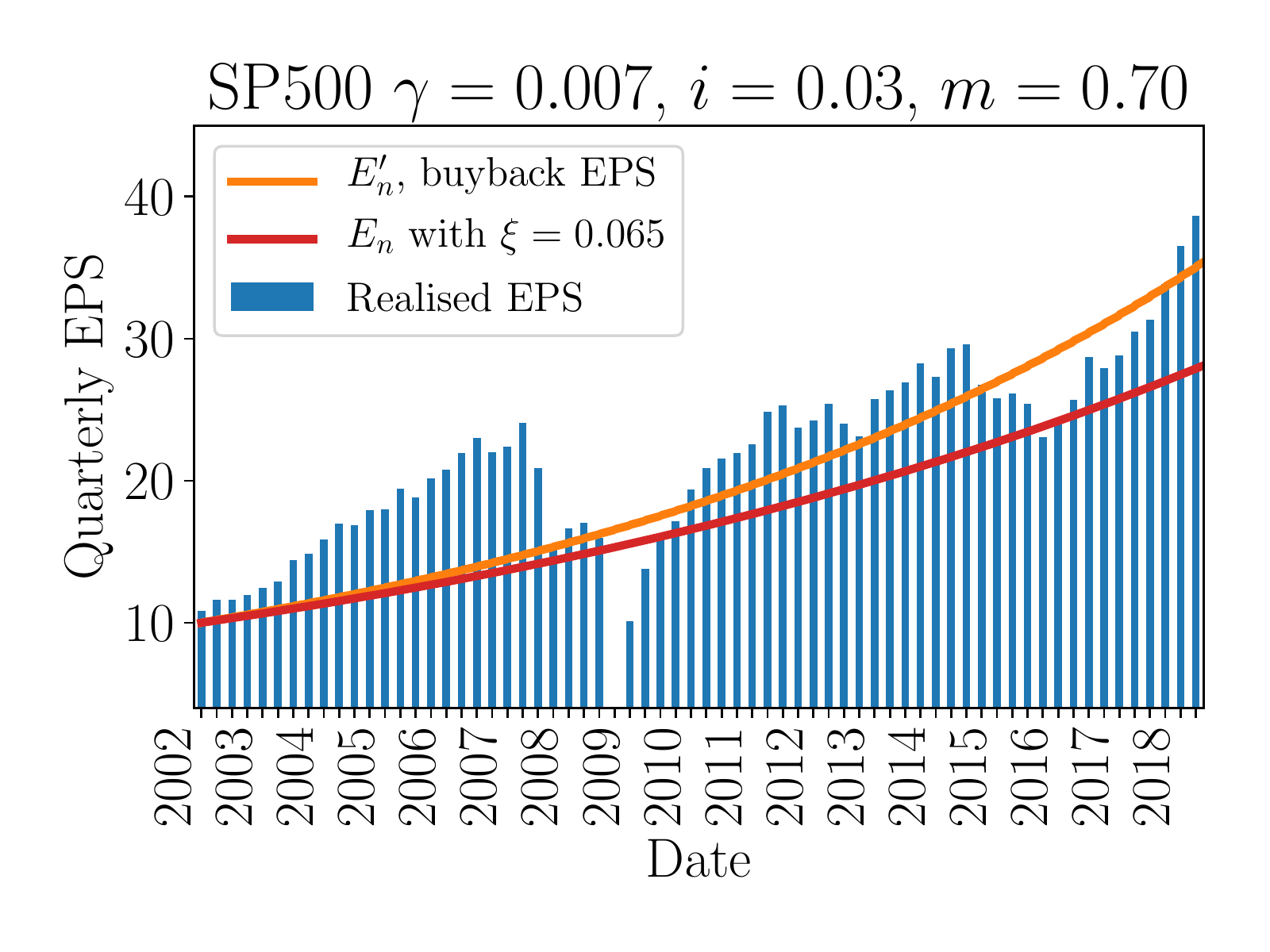}
\par\end{centering}
\caption{Earnings per share in the presence, $E_{n}'$, and absence, $E_{n}$,
of buybacks for S\&P 500 operating earnings per share (quarterly)
for the given values of $m$, $\gamma$ and $i$. We compare the growth
in earnings implied by the above analysis $E_{n}'$ (green) and contrast
this with the natural growth rate of $E_{n}$ with $\xi=0.065$ (red).}
\label{fig:spbuybacks}
\end{figure}

We consider the natural growth rate for the S\&P 500 based on the
values of the critical P/E ratio identified above from the quarter
ending in $31^{st}$ March 2002 to the quarter ending $29^{th}$ March
2018. In this case, we take values for the S\&P 500 P/E ratio, estimating
$m$ via $m=\frac{P}{P^{*}}=\frac{P_{E}}{P_{E}^{*}}=\frac{P_{E}}{i(1-t_{tax})}$.
Figure \ref{fig:spm} and plots values of $m$ given by taking the
ratio of P/E to the critical P/E, we see that over the financial crisis
of 2008/2009 the value for $m$ spikes. This is to a degree as expected
owing to the sharp dip in earnings over this time, though the interest
rate and tax rate determining $P_{E}^{*}$ is relatively constant.
As a result, we take the mean value of $m=0.7$ though note a more
refined analysis would consider the variability in natural growth
rate as well as the varying values of $m$ through time. We next plot
the values of $\gamma$ over this period in Figure \ref{fig:spgamma}
where we see the average value is $0.007$ (1 s.f.). Based on these
values we plot the values of the S\&P 500 quarterly earnings with
the compounded earnings enhancements from share buybacks overlayed
in Figure \ref{fig:spbuybacks}. We see that over the last 16 years
while the earnings per share accounting for buybacks increased by
approximately 364\%, the natural growth rate of the S\&P 500 was closer
to 287\%, suggesting that approximately 30\% of the growth (1-$\frac{187}{264}$)
over this period is as a result of share repurchases of its constituents.

\section{Conclusion}

The preceding work provides a quantitative analysis on the mechanical
effect of share repurchases on earnings per share. When shares are
repurchased for a price below their critical price, the increase in
earnings per share can be estimated from a simple cash flow argument.
When shares are repurchased as part of a buyback program we are able
to estimate the growth in earnings per share that follows as a consequence
of the program under both a geometric and arithmetic model for natural
earnings growth. We applied the methodology to the US market, analysing
the S\&P 500 for which we see that buybacks account for around 30\%
of the earnings growth over the last 16 years and over a quarter of
the growth in the case of Apple over the last 5 years.

\clearpage\bibliographystyle{plain}
\bibliography{eps_buyback}

\clearpage{}

\appendix

\section{Appendix}

\subsection{Proof of Proposition \ref{prop:The-change-in}}
\begin{proof}
We consider the earnings at time $T$ after a buyback at time $t$.
If we assume that no event affects the number of shares in the period
between $t$ and $T$ we have that, $N_{t}=N_{T}$ and in the case
of a buyback $N_{t}'=N_{t}-\frac{S_{t}}{P_{t}}$. As such we have
the earnings at time $T$ given by 
\begin{align}
E_{T}' & =\frac{\alpha\left(O_{T}+(C_{t}-S_{t})iI_{t,T}\right)}{N_{t}'}\nonumber \\
 & =\frac{\alpha\left(O_{T}+(C_{t}-S_{t})iI_{t,T}\right)}{N_{t}-\frac{S_{t}}{P_{t}}}\nonumber \\
 & =\frac{\alpha}{1-\frac{S_{t}}{P_{t}N_{t}}}\left(\frac{O_{T}+C_{t}iI_{t,T}}{N_{t}}-\frac{S_{t}iI_{t,T}}{N_{t}}\right)\nonumber \\
 & =\frac{1}{1-\gamma_{t}}\left(E_{T}-\frac{S_{t}i\alpha}{N_{t}}I_{t,T}\right)\nonumber \\
 & =\frac{1}{1-\gamma_{t}}\left(E_{T}-m_{t}\gamma_{t}I_{t,T}E_{t}\right)\label{eq:eT}
\end{align}
and using the definition of the critical P/E ratio $P_{E}^{*}(t):=\frac{P_{t}^{*}}{E_{t}}=\frac{1}{\alpha i}$.
Provided $E_{T}\ge m_{t}\gamma_{t}I_{t,T}E_{t}$ (i.e. that $m_{t}\gamma_{t}\le1$
in the presence of low interest rates and accretive buybacks) we have
that 
\begin{align*}
E_{T}' & =\frac{E_{T}-m_{t}I_{t,T}E_{t}\gamma_{t}}{1-\gamma_{t}}\\
 & =E_{T}+(E_{T}-m_{t}I_{t,T}E_{t})\sum_{n=1}^{\infty}\gamma_{t}^{n}\\
 & =E_{T}+(E_{T}-m_{t}I_{t,T}E_{t})\gamma_{t}+\mathcal{O}(\gamma_{t}^{2})\\
 & \ge E_{T}+\left(E_{T}-m_{t}I_{t,T}E_{t}\right)\gamma_{t}
\end{align*}
where the remainder is positive by inspection. The series expansion
can be derived through the identity 
\begin{align*}
\frac{a-bx}{1-x} & =\lim_{M\rightarrow\infty}(a-bx)\sum_{n=0}^{M}x^{n}\\
 & =\lim_{M\rightarrow\infty}(a\sum_{n=0}^{M}x^{n}-b\sum_{n=1}^{M+1}x^{n})\\
 & =\lim_{M\rightarrow\infty}(a+(a-b)\sum_{n=0}^{M}x^{n}-bx^{M+1})\\
 & =a+(a-b)\sum_{n=0}^{\infty}x^{n}
\end{align*}
for $|x|\le1$. 
\end{proof}

\subsection{Arithmetic earnings growth}

In the following we consider the arithmetic growth rate $E(t+\Delta t)-E_{t}=\Delta x$.
We model both the natural earnings $E_{t}$ and the accumulated interest
rate $I_{t,T}$ as stochastic processes. For the purposes of simplicity
we consider buybacks occuring at regular times $\{0,\Delta t,2\Delta t,3\Delta t,...\}$
so that we are able to consider only discrete time processes. Letting
$\mathds{E}[X]$ denote the expectation of a random variable $X$
we assume that $E_{t}$ follows some random walk with constant expected
mean, i.e. that
\[
\bar{x}:=\mathds{E}[E(t+\Delta t)-E_{t}|E_{t}]
\]
is constant for all $t$. We additionally assume the expected interest
rate over a single $\Delta t$ period is approximately constant, i.e.
$\mathds{E}\left[I(t,t+\Delta t)\right]=(1+\imath)^{\Delta t}$ ,
so from (\ref{eq:eT}) we have that the expected earnings after a
buyback occuring at time $t$ a unit of time $\Delta t$ later is
\begin{align*}
\mathds{E}\left[E'(t+\Delta t)\right] & =\mathds{E}\left[\frac{E(t+\Delta t)-m_{t}I(t,t+\Delta t)E_{t}}{1-\gamma_{t}}\right]\\
 & =\mathds{E}\left[\mathds{E}\left[\frac{\Delta x+E_{t}-m_{t}I(t,t+\Delta t)E_{t}}{1-\gamma_{t}}|\{E_{t}\}\right]\right]\\
 & =\mathds{E}\left[\mathds{E}\left[\frac{\Delta x}{1-\gamma_{t}}|\{E_{t}\}\right]+\mathds{E}\left[\frac{1-m_{t}I(t,t+\Delta t)}{1-\gamma_{t}}|\{E_{t}\}\right]E_{t}\right]\\
 & =\left(\frac{1-m_{t}(1+\imath)^{\Delta t}}{1-\gamma_{t}}\right)\mathds{E}\left[E_{t}\right]+\frac{\bar{x}}{1-\gamma_{t}}
\end{align*}
Such a result allows us to consider a sequence of buybacks occuring
at regular intervals, every $\Delta t$ time increments. In particular
we have that $\mathds{E}\left[E'(t+\Delta t)\right]=a\mathds{E}\left[E_{t}\right]+b$
for constants $a$ and $b$ that do not depend on earnings. Repeated
application of this identity leads to 
\begin{align*}
\mathds{E}\left[E(k\Delta t)\right] & =a^{k}E_{0}+b\frac{1-a^{k}}{1-a}\\
 & =\left(\frac{1-(1+\imath)m\gamma}{1-\gamma}\right)^{k}E_{0}+\frac{\bar{x}}{1-\gamma}\left(\frac{1-\left(\frac{1-(1+\imath)m\gamma}{1-\gamma}\right)^{k}}{1-\left(\frac{1-(1+\imath)m\gamma}{1-\gamma}\right)}\right)
\end{align*}
making the simplifying assumption that $\gamma_{t}=\gamma$, i.e.
the fraction spent on buybacks at any time instance is approximately
constant for all $t$. As a result we see that even when a company's
expected growth rate is close to 0 in the absence of a repurchase
program we have that
\[
\mathds{E}\left[E(k\Delta t)\right]=\left(\frac{1-(1+\imath)m\gamma}{1-\gamma}\right)^{k}E_{0}
\]
suggesting even under an arithmetic model for earnings growth, the
affect on earnings of a buyback program can not only appear to inflate
expected earning but do so also at a geometric rate. Furthermore,
we see if in addition interest rates are low then 
\[
\mathds{E}\left[E(k\Delta t)\right]\approx\left(\frac{1-m\gamma}{1-\gamma}\right)^{k}E_{0}
\]
which is precisely the $k^{th}$ power of the earnings enhancement
occuring instantaneously at $k=0$, suggesting that in such a regime
we may be able to approximate the earnings enhancement through time
by simply taking the power of the instantaneous earnings enhancement.
Indeed, it suggests even though a company's earnings may not intrinsically
be growing, if it is able to buyback its shares cheaply it is able
to appear to grow geometrically. 

\subsection{Stochastic earnings enhancement}

We here introduce a simple model for uncertainty of the above where
we make the simplifying assumption that the earnings per share cannot
be negative. In this case we are able to instead treat $i(t)\in\mathds{R}^{+}$
and $x(t)\in\mathds{R}^{+}$ as discrete time stochastic processes
defined on a probability space $(\Omega,\mathcal{F},\mathds{P})$,
with $\Omega=\mathbb{N}\times\mathds{R}^{+}$ and $\mathcal{F}$ the
generated $\sigma$-algebra. We see in this case that the earnings
per share with and without a buyback are now stochastic processes.
Following the reasoning as before, we are able to construct the following
process quantifying the relative change between the earnings with
and without a buyback as 
\[
G(t):=\frac{E_{t}'}{E_{t}}=\frac{1}{1-\gamma}\left(1-m\gamma\frac{i(t)}{x(t)}\right)
\]
For the purposes of tractability, if we let $i(t)=\prod_{p=1}^{t}(1+\epsilon_{p})$
and $x(t)=\prod_{p=1}^{t}(1+\delta_{p})$ where $1+\epsilon_{p}\stackrel{i.i.d}{\sim}\ln\mathcal{N}(\mu_{i},\sigma_{i}^{2})$
and $1+\delta_{p}\stackrel{i.i.d}{\sim}\ln\mathcal{N}(\mu_{x},\sigma_{x}^{2})$
(we assume that both are independent of each other) then we recover
the following exact expressions for the mean of the increase in earnings
as
\begin{align*}
\mathds{E}\left[G(t)\right] & =\frac{1}{1-\gamma}\left(1-m\gamma\mathds{E}\left[\frac{i(t)}{x(t)}\right]\right)\\
 & =\frac{1}{1-\gamma}\left(1-m\gamma\mathds{E}\left[\prod_{p=1}^{t}\frac{1+\epsilon_{p}}{1+\delta_{p}}\right]\right)\\
 & =\frac{1}{1-\gamma}\left(1-m\gamma\exp\left(t\left(\mu_{i}-\mu_{x}+\frac{\sigma_{i}^{2}+\sigma_{x}^{2}}{2}\right)\right)\right)
\end{align*}
using standard properties of the lognormal distribution. Such a result
suggests that as either the interest rate or growth rate volatility
increases the expected increase is dampened through time. Finally,
we see that the standard deviation of the process at time $t$ is
expressed as 

\begin{align*}
\sigma(t) & :=\sqrt{\mathds{V}\left[G(t)\right]}=\frac{m\gamma}{1-\gamma}\exp\left(\mu_{i}-\mu_{x}+\frac{\sigma_{i}^{2}+\sigma_{x}^{2}}{4}\right)\sqrt{\exp\left(\frac{\sigma_{i}^{2}+\sigma_{x}^{2}}{2}\right)-1}
\end{align*}
where we see a monotone increase with $\gamma$ (noting that $\gamma\in[0,1]$),
$\mu_{i}-\mu_{x}$, $\sigma_{i}^{2}$ or $\sigma_{x}^{2}$ .

\end{document}